\documentclass[aps,prb,preprint,showkeys,amssymb,amsmath,nobibnotes,
nofootinbib,superscriptaddress]{revtex4}

\usepackage{enumerate}
\usepackage{color}
\usepackage[colorlinks=true,linkcolor=red]{hyperref}
\usepackage{amsmath}
\usepackage{amsfonts}
\usepackage{amssymb}
\usepackage{verbatim}
\usepackage{times}
\usepackage{epic,eepic}
\usepackage{multirow}

\usepackage{graphicx}

\newcommand{\pochh}[2]{\left( #1; #2 \right)_\infty}

\newcommand{\bra}[1]{\langle #1|}
\newcommand{\ket}[1]{|#1\rangle}
\newcommand{\TrLong}[1]{{\mathrm{Tr}}\left(#1\right)}
\newcommand{\Tr}{{\rm Tr}}

\newcommand{\Ord}[1]{{\mathcal O}\!\left(#1\right)}
\newcommand{\tq}{\tilde q}
\def\cS{{\mathcal S}}

\def\ZZ{{\cal Z}}
\newcommand{\ee}{\end{equation}}
\newcommand{\be}{\begin{equation}}
\newcommand{\bea}{\begin{eqnarray}}
\newcommand{\eea}{\end{eqnarray}}

\newcommand{\eu}{{\rm e}}
\newcommand{\ii}{{\rm i}}

\newcommand{\ddelta}{\Gamma}
\newcommand{\danom}{\Delta}
\newcommand{\danomPf}{\Delta^{\text{pf}}}
\newcommand{\danomeps}[1]{\Delta^{(\epsilon)}_{#1}}
\newcommand{\GS}{0}
\newcommand{\cpf}{c_r^{\text{pf}}}
\newcommand{\rhoCTM}{\rho_{\text{CTM}}}
\newcommand{\nupf}{\nu^{\text{pf}}}

\begin{document}
\preprint{MIT-CTP 4371} 

\author{A.De Luca}
\affiliation{SISSA - 
via Bonomea 265, 34136, Trieste, Italy}
\affiliation{LPT - ENS,
24 rue Lhomond, 75005, Paris, France}
\author{F. Franchini}
\email{fabiof@mit.edu}
\affiliation{Department of Physics, Massachusetts Institute of Technology,
Cambridge, MA 02139, U.S.A.}
\affiliation{SISSA and I.N.F.N, Via Bonomea 265, 34136, Trieste, Italy}

\title{Approaching the RSOS critical points through entanglement: \\ one model for many universalities}

\begin{abstract}
We analytically compute the Renyi entropies for the RSOS models, representing a
wide class of exactly solvable models with multicritical conformal points
described by unitary minimal models and $\mathbb{Z}_n$ parafermions. The exact
expressions allow for an explicit comparison of the expansions around the
critical points with the predictions coming from field theory. In this way it
is possible to point out the nature of the so-called ``unusual corrections'',
clarifying the link with the operator content, the role of the symmetries and
the boundary conditions. By choosing different boundary conditions, we can
single out the ground states as well as certain combinations of high energy
states. We find that the {\it entanglement spectrum} is given by 
operators that are not present in the bulk Hamiltonian, although they belong to the
same representation of a Virasoro Algebra. In the parafermionic case we observe unexpected
logarithmic corrections.
\end{abstract}


\maketitle 

\section{Introduction}
Entanglement is the unique feature distinguishing a quantum system from a classical one \cite{amic2008, lato2009}. While we still lack a fundamental, general definition of what entanglement is, we can characterize it well when we consider the mutual entanglement of two complementary components of a system in a definite state, the so-called bipartite-entanglement \cite{cala2009}. A popular way to quantify it is given by the {\it entanglement entropies}. In the recent years, it has become increasingly important to be able to compute them, either numerically or analytically \cite{cira2009}. Typically, one considers the ground state $\ket{\GS}$ of a quantum Hamiltonian. Once the system has been divided into two parts $A,B$, it is possible to introduce the reduced density
matrix, tracing out one of the two
subsystems
\be
\label{rdm}
\rho_A \equiv \Tr_B \ket{\GS}\bra{\GS}
\ee
and then the Renyi entropies are defined as
\be
   {\cS}_{\alpha} \equiv {1 \over 1-\alpha} \ln \Tr \rho_A^\alpha \; ,
   \label{eq:renyi1}
\ee
We notice that, thanks to the free parameter $\alpha$, the knowledge of the Renyi entropies is equivalent to the knowledge of the full spectrum of the reduced density matrix \cite{fran2010}, whose logarithm is known as the {\it entanglement spectrum} \cite{liha2008}. A particularly important point is the $\alpha \to 1$ limit, known as Von Neumann entropy
\be
\cS = \lim_{\alpha \to 1} \cS_\alpha = - \Tr \rho_A \ln \rho_A \:,
\ee
which provides a good quantification for the entanglement in terms of a single number.

Gapped $d+1$-dimensional systems obey the so-called \textit{area-law} \cite{eise2010}: at the leading order in the thermodynamic limit of large subsystem sizes, the entanglement entropy is proportional to the area of the boundary separating $A$ and $B$. In $d=1$, such law predicts a saturation to a constant of the entanglement entropy when $A$ is composed of large intervals. For $d >1$ the area law remains true for most gapless systems, with possible logarithmic corrections \cite{swin2011}. These logarithmic contributions are a signature of $d=1$-dimensional physics. In fact, exploiting the conformal invariance of gapless $1+1$ models,
it is known that the entropy grows logarithmically with the length of the $A$ interval $\ell$, with a proportionality given by the conformal anomaly \cite{cala2004}.
In \cite{card2010}, the sub-leading contributions were analyzed and the emergence of {\it unusual corrections} was linked to the effect of relevant (and irrelevant) operators of the critical theory:
\be
   {\cal S}_\alpha = {c \over 6} \left( {1 + \alpha \over \alpha} \right) \ln \ell + c^\prime_\alpha
   + b_\alpha \ell^{-2 x/\alpha} + \ldots \; ,
   \label{SCFT}
\ee
where $c$ is the central charge of the CFT, $c^\prime_\alpha$ and $b_\alpha$ are non-universal constants and $x = \Delta + \bar{\Delta}$ is the dimension of the operator ``responsible'' for the correction. Marginal operators act differently (renormalizing the central charge) and give rise to logarithmic corrections of the form $\left( \log \ell \right)^{-2 n}$.

Similar results have been observed \cite{cala2010, erco2011, erco2012} close to
the critical points, where now the correlation length $\xi$, being large but
finite, becomes the relevant length scale (that is $\ell \gg \xi$), and have led to the conjectured form
\be
   {\cal S}_\alpha = {c \over 12} \left( {1 + \alpha \over \alpha} \right) \ln {\xi \over a_0} + A_\alpha
   + B_\alpha \xi^{-h/\alpha} + \ldots \; ,
   \label{Sexp}
\ee
where $a_0$ is a short distance cutoff, $A_\alpha$ and $B_\alpha$ are again non-universal
constants and $h$ can be interpreted as the dimension of a relevant operator\footnote{The factors of $2$ difference between \eqref{SCFT} and \eqref{Sexp} is due to the number of boundaries dividing $A$ and $B$: $2$ for an interval, $1$ in the gapped phase, where both $A$ and $B$ are taken semi-infinite.}. In general, one should not expect the entropy to be the same scaling function in $\ell$ and in $\xi$: although the coefficients of the leading term have to coincide, the operators acting in the two cases can be different and $x$ and $h$ can be different. The crossover function between the two scaling regime might be accessible using techniques similar to \cite{card2008, cast2008, doyo2009}.

Thus, while the entanglement entropy of a gapped $d=1$ system could seem not very
interesting, since it saturates to a constant in the thermodynamic limit, its
study close to a critical point could shed light on the scaling theory governing the
lattice models and its universal features. Moreover, a recently proposed
protocol \cite{card2011, aban2012} would allow the measurement of the Renyi
entropies only for gapped systems, thus rendering the theoretical computation of
the limiting value \eqref{Sexp} amenable to cold-atom experimental
confirmation.

In this paper we will focus on the analytic computation of the Renyi entropy in
this thermodynamic limit for the quantum systems obtained from a class
of integrable lattice models known as \textit{Restricted Solid-on-solid}
(RSOS)\cite{andr1984}. These models and their structures have inspired the
discussion in the last section of \cite{cala2010}, on a general relation between
the entanglement entropy of quantum (integrable) models and Virasoro characters.
Inspired by these considerations, we expand and detail the calculation sketched
in \cite{cala2010}, and extend it to the parafermionic case. The importance of
the RSOS models is multifold: first of all, they provided the first lattice
realization of the unitary conformal models \cite{bela1984} as pointed out in
\cite{huse1984, bazh1989}. In a different phase, they also realize parafermionic
models and thus give access to consistent $c >1$ CFT's \cite{zamo1985}.
Moreover, thank to the rich underlying mathematical structure, they appeared as
a 
fascinating link between integrable lattice models and number theory. 

While entanglement is associated to a quantum state, here we will
take advantage of the well-known link that allows to derive a quantum
Hamiltonian from the row-to-row transfer matrix of an integrable classical
model. In this way, the classical configuration with the lowest free energy
corresponds to the ground state of the quantum model. Moreover, by a proper
choice of the boundary conditions, one can select higher energy configurations,
which correspond to the lowest quantum state within a given sector of the
Hilbert space, and is thus a way to investigate the entanglement entropy of
states other than the ground state. Without showing explicitly the quantum
Hamiltonian associated to the RSOS transfer matrix, it is worth saying that it
naturally arises in the context of loop models \cite{fend2006}. More recently,
an explicit realization of these Hamiltonians has be obtained from a very
different perspective as a chain of interacting non-abelian anyons
\cite{treb2008}.
Another possible approach, whose terminology we decide to adopt
here, is to interpret the RSOS models as the lattice realizations of an
integrable thermal perturbation of a class of rational CFTs. This allows to
write the action as
\be
\label{perturbedAction}
\mathcal{A} = \mathcal{A}_{\mbox{\tiny CFT}} + \lambda \int d^2 x \; \epsilon (x) 
\ee
where $\epsilon(x)$ is the operator representing the thermal perturbation and $\lambda$ is the coupling constant measuring the distance from criticality.

The RSOS models being ubiquitous and integrable makes the
computation of the Renyi entropy not only interesting, but also possible
analytically by means of the \textit{Corner Transfer Matrix} (CTM) approach. In
fact, the reduced density matrix of an half-interval in the thermodynamic limit
can be shown to be equal (except for the normalization factor) to the CTM
operator \cite{pesc1999, erco2010}
\be
\label{partitionCTM}
\hat\rho_A = \ZZ_1^{-1} \rhoCTM \;, \qquad \ZZ_\alpha \equiv \Tr
\rhoCTM^\alpha \; ,
\ee
and one can therefore compute the Renyi entropy as
\be
\label{RenyiPartition}
 {\cS}_\alpha = \frac{\alpha}{\alpha-1} \ln \ZZ_1 + \frac{1}{1-\alpha} \ln
\ZZ_\alpha \; .
\ee

Even though this procedure looks similar to the replica trick exploited in the conformal case, here $\alpha$ can be an arbitrary real (or even complex) parameter, thus avoiding all the subtleties of the analytic continuation from $\alpha = n \in \mathbb{N}$, necessary to compute, for instance, the Von Neumann entropy.
Therefore, beyond checking that the conjectured form of Eq.\eqref{Sexp} applies
for the RSOS, our results provide the umpteenth check to the Cardy-Calabrese
formula \cite{cala2009-1}, both for minimal models, where the replica trick
introduces operators not present in the original Kac table, and also in a
systems with central charges greater than unity. For minimal models, we identify
the leading {\it unusual correction} of \eqref{Sexp} as coming from the second
most relevant operator in the model, that is $\danom_{3,3}$, since the most
relevant one, $\danom_{2,2}$, is odd under the $\mathbb{Z}_2$ symmetry of the
ground state. However, we find that, by varying the boundary conditions,
different sectors can be traced out and the leading correction to the entropy can
come from other operators as well. Let us stress that this result is different
from the correction one could na\"ively expect taking the anomalous dimension of the perturbation 
in \eqref{perturbedAction}.
For parafermionic models, we find the leading correction to come from the first thermal field,
 which is the most relevant only among the $\mathbb{Z}_{r-2}$ neutral fields. In addition, some
boundary conditions at infinity turn on logarithmic corrections, different from
those in \eqref{Sexp}. It would be tempting to interpret these terms
as due to a marginal operator in accordance with \cite{erik2008,
card2010}; however
these corrections are present even when the theory does not seem to support a
marginal field (which is normally related to the existence of a free boson and
present only for certain given values of $r$).
Thus, the origin of these terms still needs a full explanation and is probably
rooted in a choice of boundary conditions which has no conformal counterpart in
the continuum limit.

The paper is organized as follows: in section \ref{themodel} we introduce the RSOS models, their phase diagrams and some details about the exact solution in Regime III and I on which we will focus. In section \ref{entropyIII}, we will concentrate on the computation of the Renyi entropy for regime III, corresponding to unitary minimal models and in section \ref{entropyI} on regime I corresponding to parafermionic CFT. To better elucidate the meaning of our formulae, we will conclude the analysis with the specific examples of the Ising and 3-state Potts model in section \ref{examples}. Finally in section \ref{conclusions} we will discuss our results and their meaning. We collect some useful definitions and identities on elliptic functions in \ref{ellapp}.

\section{The model}
\label{themodel}

\subsection{Definition}

We consider the restricted solid-on-solid (RSOS) on a square lattice, first introduced in \cite{andr1984}. The variables on each node are called ``heights'' and are integer numbers restricted to the interval:
\be  1 \leq l_i \leq (r-1) \ee
A local constraints is imposed to every configuration
\be
   |l_i - l_j | = 1
   \label{constraint}
\ee
for each pair of nearest-neighbor $i,j$. The model belongs to the family of \textit{interaction round-a-face} (IRF) models, introduced by Baxter \cite{baxt1982}. Each plaquette is given a Boltzmann weight according to the configuration of the 4 sites enclosing it: $W(l_1,l_2,l_3,l_4)$. Here the four sites around the plaquette are counted clockwise from the northwest $l_1,l_2,l_3,l_4$.
\begin{figure}[h]
\begin{center}
\setlength{\unitlength}{3947sp}%
\begingroup\makeatletter\ifx\SetFigFont\undefined%
\gdef\SetFigFont#1#2#3#4#5{%
  \reset@font\fontsize{#1}{#2pt}%
  \fontfamily{#3}\fontseries{#4}\fontshape{#5}%
  \selectfont}%
\fi\endgroup%
\begin{picture}(1821,1590)(3451,-3331)
\thinlines
{\color[rgb]{0,0,0}\put(3676,-3136){\framebox(1275,1200){}}
}%
\put(3451,-1861){\makebox(0,0)[lb]{\smash{{\SetFigFont{12}{14.4}{\familydefault}{\mddefault}{\updefault}{\color[rgb]{0,0,0}$l_1$}%
}}}}
\put(4801,-1861){\makebox(0,0)[lb]{\smash{{\SetFigFont{12}{14.4}{\familydefault}{\mddefault}{\updefault}{\color[rgb]{0,0,0}$l_2$}%
}}}}
\put(4801,-3286){\makebox(0,0)[lb]{\smash{{\SetFigFont{12}{14.4}{\familydefault}{\mddefault}{\updefault}{\color[rgb]{0,0,0}$l_3$}%
}}}}
\put(3526,-3286){\makebox(0,0)[lb]{\smash{{\SetFigFont{12}{14.4}{\familydefault}{\mddefault}{\updefault}{\color[rgb]{0,0,0}$l_4$}%
}}}}
\put(3740,-2536){\makebox(0,0)[lb]{\smash{{\SetFigFont{12}{14.4}{\familydefault}{\mddefault}{\updefault}{\color[rgb]{0,0,0}$W(l_1,l_2,l_3,l_4)$}%
}}}}
\end{picture}%
\end{center}
\end{figure}
The model can be exactly solved for a proper choice of the weights $W$ where it appears as a consistent restriction of the solid-on-solid (SOS) model and hence with the same Yang-Baxter algebra of an eight-vertex model. Weights are thus parameterized in terms of elliptic functions and for the details we refer to the original work \cite{andr1984}.

At fixed maximum-height $r$, the phase-space of the model can be characterized by two parameters $p, v$.
The requirement of real and positive Boltzmann weights gives the constraints
$$ -1 < p < 1 \qquad -\eta < v < 3\eta $$
naturally arranged in four, physically distinct, regimes
\begin{eqnarray*}
\label{regimesRange}
\text{I} & -1<p<0 &\eta<v<3\eta\\
\text{II} & 0<p<1 &\eta<v<3\eta\\
\text{III} & 0<p<1&-\eta<v<\eta\\
\text{IV} & -1<p<0&-\eta<v<\eta
\end{eqnarray*}
The parameter $\eta$ is related to $p$ by
$$ \eta \equiv \frac {K(p)} r $$
where $K(p)$ is the complete elliptic integral with elliptic ``nome'' $p$.
\begin{figure}[h]
\begin{center}
\includegraphics[width=0.4\textwidth]{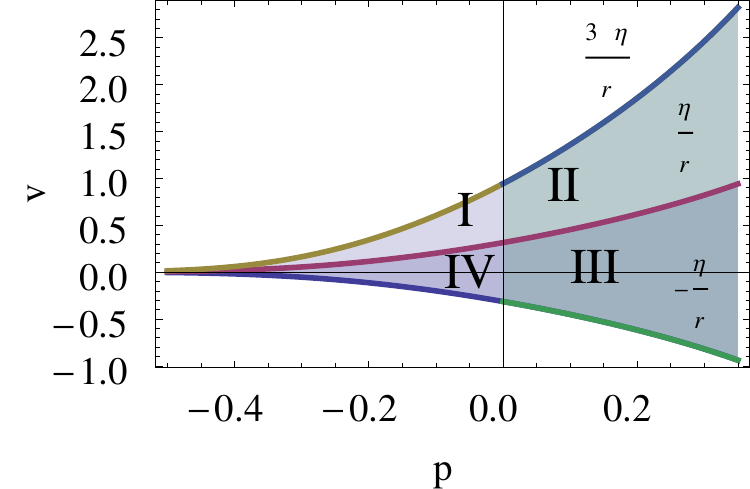}
\caption{The four regimes in the $(p, v)$ plane.}
\label{phaseDiagram}
\end{center}
\end{figure}
The regimes I, II and III, IV are separated by a line of critical points at $p = 0$.
The parameter $v$ can be considered, roughly speaking, as the spatial anisotropy of the interactions in the model and does not enter in the order parameters and the critical behavior. So, for fixed regime we will ignore it. The manifolds of exact solution will be simply lines parameterized by $p \in (-1,1)$.
By comparison with \eqref{perturbedAction}, we have that close to criticality, i.e. when $|p|\ll1$, $p \simeq \lambda$.

\subsection{Exact solution}

The exact solution in \cite{andr1984} consists of three parts.
\begin{figure}[h]
\begin{center}
\label{cornerABCD}
\vspace{1em}
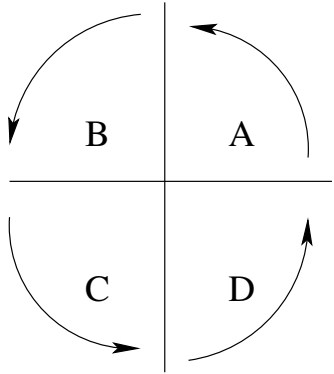
\caption{The action of the four CTM generates the full partition function.}
\end{center}
\end{figure}
\begin{enumerate}
\item  introduce the corner transfer matrix (CTM) that, once diagonalized, allows reducing the 2D
configuration sum into a 1D sum already at finite size;
\item perform the thermodynamic limit by transforming the finite size expressions into series involving gaussian
polynomials and then taking the limit as modular functions;
\item sum up the partial traces (for fixed value of the central height) of the CTM obtaining the full partition function.
\end{enumerate}
Here we will briefly review the first two steps, that are functional to our derivation. For the last step, we will use a slightly different approach with respect to the traditional one, formulated in terms of the dual variables.

\subsubsection*{Corner transfer matrix}

The method of the corner transfer matrix (CTM), introduced by Baxter \cite{baxt1982}, allows the exact solution of lattice integrable models, computing both the partition function and the one-point correlation function (e.g. the magnetization).
As shown in fig.\,\ref{cornerABCD}, four operators ${\bf A},{\bf B},{\bf C},{\bf D}$ are introduced.
${\bf A}$ is the partition function of the system restricted to the first quadrant and
with fixed boundary conditions on the positive $x$ and $y$ axis. Similarly the
other operators ${\bf B},{\bf C},{\bf D}$ are defined in the other quadrants and it follows that
\be
\label{partitionABCD}
\left(\rhoCTM\right)_{l_1\ldots l_N}^{l_1',\ldots,l_N'} =
({\bf ABCD})_{l_1\ldots l_N}^{l_1',\ldots,l_N'}\quad \Rightarrow \quad \ZZ =
\Tr{\rhoCTM}
\ee
The local height probability (LHP) for the height $l_1$ at the
origin can be written as
\be
\label{lhp}
 P_l \equiv \operatorname{Prob}(l_1 = l) = \ZZ^{-1}\TrLong{\delta_{l_1,l} {\bf ABCD}}
\ee
The CTM formalism becomes particularly useful in integrable lattice models,
where it becomes possible to fully diagonalize the operator
$\rhoCTM$, hence computing the exact spectrum and, thus, its
trace. In the  RSOS case, the last two sites $m+1, m+2 \equiv N$, determine the
boundary conditions and we will take them as fixed. Once in the eigenbasis, the
corresponding diagonal operator $\rho_\text{diag}$ can be decomposed as
\be
  \rho_\text{diag} = {\bf R T}
  \label{rhodiag}
\ee
where both ${\bf R},{\bf T}$ are diagonal, but ${\bf R}$ is a weight that depends only on the height at the origin $l_1$, while ${\bf T}$ takes into account the configuration on the whole line $l = \{l_1,\ldots,l_m,l_{m+1},l_{m+2}\}$. They can be
summarized in the four regimes as follows
\begin{center}
  \begin{tabular}{|c|c|c|c|c|}
    \hline
    Regime	&$t$		&$\ln x$			&$R_{l,l}$		&$T_{l,l}$\\
    \hline
    II		&$2 - r$	&\multirow{2}{*}{$\frac{4\pi^2}{r\ln|p|}$}	&\multirow{2}{*}{$x^{(2-t)(2l_1 - r)^2/16r} E(x^{l_1}, x^r)$}&	\multirow{2}{*}{$x^{t \phi \left[ l \right]}$}	\\ \cline{1-2}
    III 		&$2$	&						&		&
				\\
    \hline
    I		&$2-r$		&\multirow{2}{*}{$\frac{2\pi^2}{r\ln|p|}$}	&\multirow{2}{*}{$x^{1/4 + (2-t)(2l_1 - r)^2/8r} E(x^{l_1}, x^{-r/2})$}& \multirow{2}{*}{$x^{t \psi \left[ l \right] }$}
				\\ \cline{1-2}
    IV		&$2$		&	& &
				\\
    \hline
  \end{tabular}
  \label{RTtable}
\end{center}
where $E(z,x)$ is the elliptic function defined in \eqref{Edef} and we introduced the \textit{CTM Hamiltonians}:
\bea
 \phi \left[ l \right] &= &
 \sum_{j=1}^m j\frac{|l_{j+2} - l_j|}{4} \; ,
 \label{cornerham1}\\
 \psi \left[ l \right] &= &
 \sum_{j=1}^m j \delta_{l_j,l_{j+2}} \left\{ \delta_{l_{j+1},l_j+1} \theta \left( l_j - {r \over 2} \right)
 + \delta_{l_{j+1},l_j-1} \left[ 1 - \theta \left( l_j - {r \over 2}  \right) \right] \right\} \; ,
 \nonumber \\
\label{cornerham}
\eea
where $\theta(x)$ is the step-function with $\theta (x \le0 )=0$ and $\theta(x>0) = 1$.

\subsection{Groundstate structure and critical points}
\label{sec_groundstate}

The two functions (\ref{cornerham1},\ref{cornerham}), can be considered as Hamiltonians related to the CTM. In fact, they appear as energies for the 1d configurations in the trace sum of \eqref{partitionABCD}.
We can therefore use such expressions to deduce the form of the groundstate in
each regime, as the configurations having the maximum contribution in the trace:
since each $2d$ groundstate is invariant under a southwest to northeast translation, it
will be enough to fix it on a line $l_1, \dots, l_N$. Moreover, the constraint \eqref{constraint} naturally divides the system into two sub-lattices, one with even heights and one with odd ones. By specifying boundary conditions (at infinity and at the origin, in a consistent way), we assign a given parity to each sub-lattice.  But a translation of the whole system by a lattice site gives an equivalent configuration, with opposite parity. Thus, we can take the central height $l_1$ and use its parity to classify each ground state out of this trivial $Z_2$ degeneracy.

In the different regimes we have the following structures\footnote{To avoid additional spurious degeneracies, in this classification we will assume $r$ to be odd for regimes I and IV.}:
\begin{enumerate}[I: ]
 \item There is only one groundstate per each parity of $l_1$: $(l_i, l_{i+1}) = (n,n+1)$ and $(l_i, l_{i+1}) = (n+1,n)$ with $n \equiv \frac{r-1}{2}$.
 \item There are $2r-4$ groundstates ($r-2$ for each parity) of the form of ascending and descending sequence from $1$ to $r-1$: e.g. $(l_1,l_2,\ldots, l_r, l_{r-1}, l_r, \ldots, l_{2r-4}, l_{2r-3}) = (1,2,\ldots, r-1, r-2, \ldots, 2,1)$ and all its translated.
 \item Also in this case we have $2r-4$ groundstates, where all the odd/even sites have the same height: $l_{2i} = X$, $l_{2i+1} = Y$ with $|X-Y| = 1$.
 \item As for regime III we have a groundstate for each couple of available nearest-neighbor values except for the regime I groundstate values: thus $2r - 4 - 1 - 1 = 2r -6 $ groundstates.
\end{enumerate}
It is clear that if there is only one groundstate (for each $l_1$ parity), then
we expect the system to be ``disordered'' and this is true in regime I. Indeed,
here the order parameter is independent from the boundary conditions, within a given parity of sub-lattices.
When there is more than one ground state (per parity), the system is in
an ``ordered'' phase.

The critical points can be understood and identified with an appropriate conformal point \cite{huse1984}:
\begin{itemize}
 \item I $\leftrightarrow$ II critical point: the system passes from a disordered to an ordered phase, where $p$ acts like a temperature; the critical point has the conformal structure of parafermion.
 \item III $\leftrightarrow$ IV critical point: here both phases are ordered and the groundstate degeneracy passes from $r-2$ to $r-3$; the critical point has the conformal structure of $(r-1)$-unitary minimal model.
\end{itemize}

\subsection{Thermodynamic limit}

It is easier to approach the critical points starting from region III and I, thus, from now on, we will focus just on these  regimes. We are interested in the thermodynamic limit of the replicated partition function, introduced in \eqref{partitionCTM}. The details of the calculation for $\alpha =1$ can be found in the original work \cite{andr1984}, thus here we can concentrate only on the main points and the few modifications needed. For convenience, we collect some definitions and the relevant properties of elliptic functions in \ref{ellapp}.

\subsubsection{Regime III}

The finite-size partition function for $\alpha$ replicas is easily obtained from \eqref{rhodiag}, by summing over the value of the central height
\be
\label{partitionIII}
 \ZZ_\alpha = \sum_{1\leq a < r} \left[ E(x^{a}, x^r) \right]^\alpha X_m(a, b, c; x^{2\alpha}) \; ,
\ee
where we singled out the boundary conditions as $a = l_1, b = l_{m+1}, c = l_{m+2}$ and
$$ X_m(a,b,c;q) \equiv \sum_{l_2, \ldots, l_m} q^{\phi \left[ l \right]} \; . $$
The thermodynamic limit can be computed exactly once this expression is
rewritten in terms of gaussian polynomials \cite{andr1984}, resulting in
\be
   X (a,b,c;q) \equiv \lim_{m \to \infty} X_m(a,b,c; q) = (q)_\infty^{-1} \; q^{bc/4} \;
   \ddelta\left(a, \frac{b+c-1}{2}; q\right) \; ,
   \label{Xdef}
\ee
where the \textit{q-Pochhammer} symbol $(q)_\infty$ is defined in \eqref{qpochh} and
\bea
   \ddelta(a,d; q) & \equiv &
    q^{\frac{a(a-1)}{4}}\left\{q^{-\frac{ad}{2}} E[-q^{(r-a)(r-1) + rd}, q^{2r(r-1)}]
    \right. \nonumber \\
    && \qquad \quad \left.
    - q^{\frac{ad}{2}} E[-q^{(r+a)(r-1) + rd}, q^{2r(r-1)}]\right\} \; .
\eea
The partition function can thus be written as
\bea
 \ZZ_\alpha & = & \lim_{m \to \infty} \sum_{1\leq a < r} \left[ E(x^{a}, x^r) \right]^\alpha X_m(a, b, c; x^{2\alpha})
 \nonumber \\
 & = & x^{\frac{\alpha bc}{2}}(x^{2\alpha})^{-1}_\infty\sum_{1\leq a < r}
 \left[ E(x^{a}, x^r) \right]^\alpha \ddelta\left(a, \frac{b + c - 1}{2}; x^{2\alpha}\right) .
 \label{ZalphaIII}
\eea

\subsubsection{Regime I}

In this regime, using the table in section \ref{RTtable}, the finite-size, $\alpha$-replicated partition function is given by
\be
\label{partitionI}
\ZZ_\alpha = \sum_{1 \leq a < r} x^{a\alpha(1+a-r)/2} \left[ E(x^a, -x^{r/2}) \right]^\alpha Y_m(a,b,c; x^{\alpha(r-2)})
\ee
where we introduced
\be
\label{Yfinite}
Y_m(a,b,c; q) \equiv \sum_{l_2,\ldots, l_m} q^{\psi(l)}
\ee
As before the thermodynamic limit is computed taking the limit $m \to \infty$ in
\eqref{Yfinite}. Unlike the regime III, here the sum is not convergent, due to
the non-zero energy density of the groundstate. Thus, we factor out the diverging contribution (which amounts to an irrelevant redefinition of the partition function normalization), obtaining
\be
\label{thermoI}
\lim_{m\to \infty} q^{-m} Y_m(a,b,c; q) =  (q)^{-1}_\infty \; f_{b,c}(q) \;  E(q^a, q^r) \: ,
\ee
where the boundary conditions enter only in the function $f_{b,c}(q)$, defined as
\bea
\label{boundaryI}
&f_{b,b+1} (q) &= \left\{\begin{array}{lcl}
                        \varepsilon_{b}(q) & \qquad & 1 \leq b < n \; ,  \\
		        1 & \qquad & n \leq b \leq r-2 \,
		\end{array}\right.
\\
\label{boundaryIb}
&f_{b,b-1} (q) &= \left\{\begin{array}{lcl}
			1 & \quad & 2 \leq b \leq n+1\; \\
                        \varepsilon_{r-b}(q) & \quad & n+1 < n \le r-1 \;
		\end{array}\right.
\eea
where $n$ is the integer part of $r/2$ and we defined
$$
\varepsilon_b(q) \equiv \frac{q^{1-b} (1-q^b)}{1-q} \; .
$$

\section{Entropy and partition function: Regime III}
\label{entropyIII}

Now that we have introduced the model and the replicated partition functions, it is straightforward to proceed with the calculation of the Renyi entropy, using \eqref{RenyiPartition}.
However, before we take on the full computation, following \cite{cala2010}, we would like to exploit the known relation between the partition function with fixed boundary conditions (both at infinity and at the origin) and the characters of primary fields in minimal models \cite{difr1997}. This link will drive the expansion of the entropy around the critical point, as we will show in section \ref{centralspinIII}.
Let us discuss this point in some detail.

\subsection{Characters of the minimal models}
\label{minchar}

It was noticed in \cite{date1986, date1987, date1987-1, sale1989, behr2001} that the quantity in (\ref{Xdef})
can be identified with a minimal model character. This can be shown by simply rewriting \eqref{Xdef}, using the sum expansion in \eqref{Edef}, yielding
\be
   X (a,b,c;q) = 
   q^{\frac{1}{4}(a-d)(a-d-1) + \frac{c_r}{24} - \danom_{d,a}} \;
   \chi^{(r-1)}_{d,a} (q) \;  ,
   \label{Xchi}
\ee
which is the character in the minimal model $(r,r-1)$ of the primary with conformal dimension
\be
   \label{conformalDim}
   \danom_{d,a} = { \left[ d \, r - a (r-1) \right]^2 -1 \over 4 r (r-1)} \; ,
\ee
and central charge
\be
    {\rm c}_r = 1 - {6 \over r (r-1)} \; .
    \label{centralc}
\ee
Here the boundary conditions at infinity are accounted just by the combination
$$d \equiv {b +c -1 \over 2} \; ,$$
Therefore, $X(a,b,c; q)$ is equivalent to a generating function of the Verma module degeneracy for the representation fixed by the boundary conditions $a,b,c$.
We stress here that this equality is valid only at a formal level: indeed, in the one-dimensional configurational sums, the elliptic nome $q$ is a
measure of the departure from criticality while in the conformal characters $q$ is the modular
nome related to the geometry on the torus at criticality. Even though they are both usually
denoted by $q$, these are two very different objects.

Similarly, one should not confuse the formal identification of fields in the
Renyi entropy expansion, with the operator actually responsible for opening the
gap. We recall that regime III can be described as \eqref{perturbedAction}, that
is as a lattice deformation of a minimal model by means of a perturbation given
by $\epsilon(x) \simeq \phi_{1,3}(x)$, which is known to be both thermal and
integrable \cite{delfino2000correlators}. As we shall see, this operator does
not appear among the most relevant ones in the Renyi expansion in regime III.

\subsection{Fixed central height}
\label{centralspinIII}

We can now compute the Renyi entropy in a sector where the height at the origin is kept fixed. As already pointed out in \cite{cala2010}, we stress that this is
done at the level of the corner transfer matrix, and so, of the reduced density matrix: if such degree of freedom was fixed at the level of the Hamiltonian,
it would indeed affect the interaction between the two parts of the system.
Rather, fixing the height at the origin of the CTM corresponds to selecting a
sector out of the whole Hilbert space of the model, and taking the groundstate
within this projection. Thus, we are measuring the entanglement of the lowest
energy state within this subspace. In general, these states will be a
superposition of high energy states and thus the calculation of their Renyi
entropy can shed some light on their properties.

In approaching the gapless point, the elliptic nome $q = x^{2 \alpha}$, in (\ref{ZalphaIII}), tends to
unity. As this is not the best parameterization to extract the leading
contributions, we perform the customary dual transformation of the elliptic nome, granting us an expansion in the original parameter $p$, which tends to zero at criticality.
Using \eqref{jacobiIdentities} and \eqref{EwithTheta} for one term in the sum of
Eq. \eqref{partitionIII}, we obtain
\be
  \label{fixedPartitionFunction}
  \ZZ_\alpha^{(a)} \equiv \frac{\theta_3\left(\frac{\pi d}{2r - 2} - \frac{\pi
a}{2 r} , p^{\frac{1}{8\alpha(r-1)}}\right)-\theta_3\left(\frac{\pi d}{2r - 2} +
\frac{\pi a}{2 r}, p^{\frac{1}{8\alpha(r-1)}}\right)}{\theta_4\left  (\frac{-
\ii r \ln p}{8 \alpha },p^{\frac{3r}{4 \alpha }}\right) p^{\frac{r}{48
\alpha}} \sqrt{2r (r-1)}} \; ,
\ee
where the index $a$ refers to the fixed value of the height at the origin. As
stated in \cite{cala2010}, the correspondence with a conformal character,
allows to reinterpret this duality, at the very end grounded on the Poisson resummation formula, as the invariance of the torus under the modular group. In a CFT, every character can then be expressed as a linear combinations of characters of the dual theory \cite{difr1997}
\be
   \chi_{t,s}^{(r-1)}  \left( \tilde{q}\equiv \eu^{- \ii \pi/\tau} \right) = \sum_{t',s'} S_{t,s}^{t',s'} \chi_{t',s'}^{(r-1)} \left( q = \eu^{\ii \pi \tau} \right) \; ,
   \label{dualchi}
\ee
where
\begin{equation}
  \label{modularmatrix}
  S_{t,s}^{t',s'} = 2 \sqrt{\frac{2}{r(r-1)}} (-1)^{(t+s)(t'+s')}
  \sin \left( \pi \frac{t t'}{r-1}\right) \sin \left(\pi  \frac{s s'}{r} \right)
\end{equation}
is the so-called {\it modular matrix}.

To reproduce this result in our setting, we can expand \eqref{fixedPartitionFunction} using \eqref{thetadef} and
\be
\label{logexpansion}
 \ln \left(q\right)^{-1}_\infty
    = \sum_{n=1}^\infty \sum_{k=1}^\infty \frac{q^{n k}}{k}
   = \sum_{n=1}^\infty \sigma_{-1} (n) q^{n}    =  q + \frac{3}{2} q^{2}
  + \frac{4}{3} q^{3} + \frac{7}{4} q^{4}
  + \Ord{q^{10}} ,
\ee
where $\sigma_{\kappa} (n)$ is the sum of the $\kappa$-th powers of the divisors of $n$
\be
\label{divisorDef}
   \sigma_{\kappa}(n) \equiv \sum_{\substack{ j < i=1 \\ j \cdot i =n}}^\infty (j^\kappa + i^\kappa)
   + \sum_{\substack{ j=1 \\ j^2 =n}}^\infty i^\kappa \; .
\ee

To compare the expansion of \eqref{fixedPartitionFunction} and \eqref{dualchi}, it is convenient to use the parameter truly dual to the one used in \eqref{Xchi}, that is
\be
\label{tqdef}
\tq \equiv p^{r \over 2} = p^{2\nu}
\ee
where $\nu =(2 - 2\danom_{1,3})^{-1} = r/4 $ is the correlation length critical exponent in Regime III
\cite{pear1998}. Collecting everything we obtain
\be
  \ln \ZZ_\alpha^{(a)} = - \frac{c_r}{24 \alpha} \ln \tq + C'_{adr}
  +  4 \gamma_{adr} \tq^{3 \over {4 \alpha r (r-1)}} -8 \gamma_{adr}^2 \tq^{3 \over {2\alpha r (r-1)}}
  + \Ord{\tq^{2 \over {\alpha r (r-1)}}} \; ,
\ee
where
\be
   \gamma_{adr} \equiv \cos\left(\frac{\pi d}{r-1}\right) \cos\left(\frac{\pi a}{r}\right)
   \label{gammadef}
\ee
and
\be
   C'_{adr} \equiv \ln \left( {4 \over \sqrt{2r(r-1)}} \sin {\pi d \over r-1} \sin{\pi a \over r } \right) \; .
   \label{Cprimedef}
\ee
is the zeroth-order correction, corresponding to the boundary entropy of \cite{affl1991}.

Using \eqref{RenyiPartition}, we can obtain the expansion for the Renyi entropy in $\tq$, while still keeping the central height fixed:
\bea
   \cS_\alpha^{(a)} & = & -\frac{c_r}{24}\left(1 + \frac{1}{\alpha}\right) \ln \tq +
   C'_{adr} +  \frac{4 \gamma_{adr}}{1 - \alpha} \left( \tq^{3 \over {4 \alpha r (r-1)}} - \alpha \tq^{3 \over {4 r (r-1)}} \right)
   \nonumber \\
   && \qquad - {8 \gamma_{adr}^2 \over 1 - \alpha} \left( \tq^{3 \over {2 \alpha r (r-1)}} - \alpha \tq^{3 \over {2r (r-1)}} \right) +
   \Ord{\tq^{2 \over {r (r-1)}}} \; .
 \label{RentropyFixed}
\eea
It is well known \cite{cala2004} that one can read off the central charge of the model from the coefficient of leading term of the entropy, as in \eqref{RentropyFixed}. Let us remark, however, that it is a pleasant check to notice that the standard conformal result, obtained using the replica trick, remains valid also for the minimal models, where the twist operator introduced in the computation does not belong to the Kac table of the CFT.

As suggested in \cite{cala2010}, the sub-leading corrections contain information on the operatorial content of the theory and their characters. In fact, from \eqref{RentropyFixed} and comparing \eqref{Cprimedef} with \eqref{modularmatrix} we recognize, consistently with \cite{cala2010},
\be
  C'_{adr} = \ln S_{1,1}^{d,a} \; .
\ee
Indeed, the zero-order term is related to the modular matrix between the primary
field chosen by the boundary condition and the identity, which is giving the
dominant contribution. The first correction in (\ref{RentropyFixed}) is coming, as expected, from the
most relevant field. Indeed, we see that:
$$ \gamma_{adr} = \frac{S_{2,2}^{d,a}}{4 S_{1,1}^{d,a}} $$
and, coherently, from \eqref{conformalDim} we recognize that the exponent of the correction is $ \danom_{2,2} = \frac{3}{4 r (r-1)} $.
The identification with the operators of a Virasoro algebra can continue to higher orders, but one should notice that the expansion of the logarithm generates additional terms which do not appear in the Kac table, such as the second sub-dominant correction in \eqref{RentropyFixed}, which is just a $2 \danom_{2,2}$. This correction is always dominant over the $\danom_{3,3} = \frac{2}{r(r-1)}$.

It should be noted here, that the parameter $\tq$ is microscopical in nature and the entropy is usually measured as a function of a thermodynamical parameter, such as the correlation length $\xi$. From \cite{Obri1997} we know that
\begin{equation}
   \xi = - {1 \over \ln k' (|p|^\nu)} = - {1 \over \ln k'(|\tq|^{1 \over 2})} \;,
   \label{xidef}
\end{equation}
where
\begin{equation}
   k' (q) = \prod_{n=1}^\infty \left( {1 - q^{2n-1} \over 1 + q^{2n-1} } \right)^4 =
   {\theta^2_4 (0) \over \theta^2_3 (0)} \; ,
\end{equation}
From these expressions we get the expansion:
\be
  \tq = \frac{1}{64 \xi^2}-\frac{1}{1536 \xi^4}+\frac{113}{2949120 \xi^6}+ \Ord{\xi^{-8}} \; ,
  \label{tqxiexp}
\ee
which should be substituted order by order in \eqref{RentropyFixed}. At the leading order, this substitution correctly fixes the usual normalization in front of the leading logarithm in terms of the central charge $c_r$ and the exponents of the corrections as $h = 2 \Delta$ in \eqref{Sexp}. The rest of the terms, however, which strictly vanish in the scaling limit, spoil the possibility of reading and reconstructing the operator content of the characters appearing in the entropy in any study at finite lattice spacing. This effect is completely analogous to the one discussed in \cite{erco2012} for the $XYZ$ chain.

\subsection{Full entropy}
\label{fullentropyIII}

To calculate the bipartite Renyi entropy of the model in its true ground state we should sum over the central height. Using the dual transformation in the  full partition function \eqref{partitionIII}, we have
\be
 \label{partitionFunction}
 \ZZ_\alpha =  \sum_{a = 1}^{r-1} \theta_1^\alpha \left(\frac{a \pi}{r}, \sqrt{p}\right) \ZZ_\alpha^{(a)} \; ,
\ee
where $\ZZ_\alpha^{(a)}$ is given by \eqref{fixedPartitionFunction}. We remark that, while $\ZZ_\alpha^{(a)}$ has a simple interpretation in terms of a character, the coefficients in the sum over the central height in \eqref{partitionFunction} do not. In the previous section, since $a$ was kept fixed, the value of the coefficient could be absorbed in the normalization of the partition function, but now we cannot ignore these contributions anymore.

For integer values of $\alpha$ this expression can be handled using the infinite sum representation of $\theta_1$ in \eqref{theta1}, giving\footnote{For arbitrary values of $\alpha$, we find the infinite product representation of the $\theta$-functions to be more convenient, although completely equivalent.}
\bea
    \ln \ZZ_n & = &  - {r  \over 48 n} \; c_r \ln p + {n \over 8} \ln p + \ln { 2^{2 +n} \over \sqrt{2 r (r-1)}}
    - \sum_{j=1}^\infty \ln \left( 1 - p^{ r \, j \over 2 n} \right)
    \label{Zalpha2} \\
    && + \ln \sum_{j=1}^\infty p^{j^2 -1 \over 8 n (r -1)} \sin \left( { \pi d j \over r-1} \right)
    \sum_{a=1}^{r-1} \sin {\pi a j \over r} \left[ \sum_{k=0}^\infty (-1)^k
    p^{k(k+1) \over 2} \sin \left( (2k +1) {a \pi \over r} \right) \right]^n \; .
    \nonumber
\eea
Take, for instance, $\alpha =1$: the sum over $a$ can be computed immediately
using the orthogonality condition
\be
    \sum_{a=1}^{r-1} \sin {\pi a n \over r} \sin  {a m\pi \over r} = {r \over 2} \: \delta_{n,m}
    \label{sumdelta}
\ee
recovering the partition function of the RSOS model with fixed boundary conditions at infinity \cite{andr1984} already in the dual formulation, that once expressed in $\tq = p^{\frac r 2}$ gives
\be
   \ZZ_1 = \sqrt {2 r \over r -1} {\theta_1 \left( { \pi d \over r-1}, \tq^{1 \over (r-1)} \right)
    \over \tq^{1 \over 24} \prod_{j=1}^\infty \left(1 - \tq^j \right) }\; ,
    \label{Z1}
\ee
For general values of $\alpha$, at our knowledge the sum in \eqref{partitionFunction} can not be computed analytically. However, it is possible to obtain its expansion order by order close to the critical point introducing the coefficients
$$ s_\alpha(n,k) \equiv \sum_{a=1}^{r-1} \sin^\alpha {\pi a \over r} \sin {\pi a n \over r} \cos^k {2 \pi a \over r} \; , $$
The first few terms give
\begin{eqnarray}
   \cS_\alpha & = &
   {c_r \over 24} \; {1 + \alpha \over \alpha} \ln \tq
   + \ln \left[ \sqrt{2 r \over r-1} \sin{ \pi d \over r-1} \right] + {1 \over 1 -\alpha} \ln \left[ {2 \over r} s_\alpha (1,0) \right]
   \nonumber \\
   && - {1 \over 1 -\alpha}  {s_\alpha (3,0) \over s_\alpha (1,0)}  \left( 4
\cos^2 {\pi d \over r-1} - 1\right) \tq^{2 \over \alpha r (r-1)}
   + \Ord{\tq^{4 \over \alpha r(r-1)}} \; ,
   \label{RentropySum}
\end{eqnarray}
Some observations about this expression are in order
\begin{itemize}
 \item it remains finite as it should, in the $\alpha \to 1$ limit due to the properties of $s_{\alpha\to 1}(n,k)$: e.g. $s_1 (3,0) = 0, s_1(1,0) = \frac r 2$;
note that this is a different mechanism w.r.t. \eqref{RentropyFixed}, where terms with and without $\alpha$ at the exponent appear in pairs and together render the Von Neumann limit finite;
 \item the leading term remains the same as \eqref{RentropyFixed} being dictated by the CFT central charge;
 \item since $ s_\alpha(2 n, k) = 0 $
for all integers $n,k$, every correction coming from the operator
$\danom_{2,2}$ and its descendants disappear and the first sub-leading term is
now related to the primary field of dimension $\danom_{3,3}$. We interpret this cancellation as due to
the $\mathbb{Z}_2$ symmetry
$$ l \to r - l $$
under which the full partition function \eqref{partitionFunction} is invariant, while the most relevant field, being identifiable with the order parameter \cite{zamo1986}, is indeed odd. Of course, this implies that all odd-operators identically vanish in the expansion of the entropy.
In any case, as we already pointed out, these corrections are not directly
ascribable to the operator opening the gap in \eqref{perturbedAction}, since in
general $\danom_{1,3} \neq \danom_{2,2},\danom_{3,3}$. This will not
be the case in regime I: as we are going to show in the next section, in the
disordered phase the leading correction seems to be given by the same operator
opening the gap.
\end{itemize}

\section{Regime I}
\label{entropyI}

We can now turn back to regime I and its bipartite entanglement entropy. As we saw in section \ref{sec_groundstate}, this regime corresponds to a disordered phase, where local expectation values are independent from the boundary conditions.
Indeed, the structure of the entanglement entropy is different from before.
As can be seen from (\ref{partitionI}, \ref{thermoI}), the contribution to the partition functions of boundary conditions at infinity factorizes out in the term $f_{b,c} (q)$. However, this contribution does not cancel out in the entanglement entropy \eqref{RenyiPartition} and can bring a finite and interesting contribution. In the analysis, we will separate the bulk and boundary contribution and consider them separately:
\be
   S_\alpha = S_\alpha^{\rm (bulk)} + S_\alpha^{(bc)} \; .
\ee
Moreover, looking at \eqref{thermoI}, we notice that, due to the $\mathbb{Z}_2$ symmetry, the fixed central height partitions function for $a$ and $r-a$ are equal and additional relations can be established for certain values of $r$ and $a$ for their coefficients in \eqref{partitionI}.

In approaching the transition toward regime IV the system undergoes a second order phase transition described by the parafermionic conformal field theory \cite{zamo1985}. Here we summarize the main features of these conformal points.

\subsection{Conformal content of parafermions}

The critical point can be described as the coset
$$\frac{\hat{sl}(2)_{r-2}}{\hat u(1)} \; ,$$
with central charge
\be
\label{centralchargeI}
 \cpf = \frac{2(r-3)}{r}\; .
\ee
Beyond the conformal one, these theories enjoy an additional
$\mathbb{Z}_{r-2} \times \tilde{\mathbb{Z}}_{r-2}$ symmetry (which is actually
enlarged to a $W_{r-2}$). This structure allows to reduce the number of allowed
anomalous dimensions to a finite set, even for $r>6$, i.e. $c>1$. These
dimensions are determined by the charges $(Q, \tilde Q)$ under the
two $\mathbb{Z}_{r-2}$ symmetries, since each of them is defined modulus $r-2$.
Following \cite{merc2001}, we introduce the two indexes
\bea
\label{ParaIndexes}
l &=&  Q - \tilde Q \; , \\\nonumber
m &=& Q + \tilde Q \; ,
\eea
in terms of which the conformal dimension of the most relevant field in each sector can be parameterized as
\be
\label{dimensionsI}
 \danomPf_{l,m} = \frac{l(l+2)}{4r} - \frac{m^2}{4(r-2)} \; , \qquad
\begin{array}{ll}
                                                   &0\leq l \leq r-2 \; ,\\
						   &0 \leq |m| \leq l \; ,\\
						  &l-m = 0\operatorname{mod} 2 \; .
                                                  \end{array}
\ee

As a matter of fact, each combination
$(\danomPf_{l,m},\bar{\Delta}^{\text{pf}}_{l',m'})$ of dimensions for the
holomorphic and anti-holomorphic part can correspond to more than one primary
field. To resolve this degeneracy, one need to look into their representation
under the $W$-algebra \cite{fate2009}. In particular, within the sector neutral
under the two $\mathbb{Z}_{r-2}$'s, i.e. with $(Q,\tilde{Q})=(0,0)$, we have
the
following allowed dimensions
\be
\label{epsilonk}
\danomeps{k} = \frac{k(k+1)}{r} \; .
\ee
These fields $\epsilon_k$, often called {\it energy} or {\it thermal fields},
are spin-less, that is $\Delta^{(\epsilon)}_k =
\overline{\Delta}^{(\epsilon)}_k$ and the identity is $\epsilon_0$. We
recognize that they are degenerate with the parafermionic operators with
$(l,m)=(2k,0)$, i.e. $\danomPf_{l = 2k,0} = \danomeps{k}$.

In regime III we showed the exact mapping existing between the partition function at fixed boundary conditions and a conformal character, see \eqref{Xchi}.
A similar relation can be established regime I as well, but it is less explicit since the mapping is no more one-to-one: we refer to \cite{merc2001} for the precise construction.

In passing, let us point out that the transition between regime I and II can be
described as \eqref{perturbedAction} where the gap-opening perturbation is due
to the most relevant thermal field $\epsilon_1 (x)$.

\subsection{Fixed central height}
\label{centralspinI}

In order to extract the behavior around criticality, we express each term of the sum in \eqref{partitionI} using the parameter $p$, through a duality transformation, as we did in \eqref{fixedPartitionFunction} and \eqref{partitionFunction}. We recall that in this regime $-1<p<0$. However, following \cite{andr1984}, the formulae for the partition function and the entropy are to be understood as depending only on the absolute value of $p$. Thus, in the following we will intend the substitution
$$ p \longrightarrow |p| = -p \; .$$
The partition function at fixed height at the origin can be written as
\be
\label{ZaI}
\ZZ_\alpha^{(a)} = \frac{\theta_1\left(\frac{a\pi}{r},p^{\frac{1}{\alpha(r-2)}}\right)}{ \theta_4\left(\frac{\ii r \log p}{2\alpha(r-2)},p^{\frac{3r}{\alpha (r-2)}}\right)
p^\frac{r}{12 \alpha(r-2)} \sqrt{r}}  \; ,
\ee
where, for the moment, we dropped the term $f_{bc}(q)$, as discussed.

The computation of the Renyi entropy is quite similar to what we did in section \ref{centralspinIII}.
Introducing again $\tq$ following \eqref{tqdef}, with $\nupf = \left(2 - 2\danomeps{1}\right)^{-1} = \frac{r}{2(r-2)} $
in this regime, we obtain
\be
\label{RentropyFixed1}
 S_\alpha^{(a)} = - \frac{\cpf}{24} \left(1 + \frac 1 \alpha\right) \log \tq
+ C'_a -  \frac{\gamma_a}{1-\alpha} \left(\tq^{\frac{2}{\alpha r}} -
\alpha \tq^{\frac{2}{r}}\right) \; ,
\ee
where we defined
\bea
 \gamma_a & = &  1 + \cos \frac{2 \pi a}{r} \; , \\
 C'_a & = & \log \left(\frac{2 \sin \frac{\pi a}{r}}{\sqrt r}\right) \: .
\eea
We recognize that, regardless of the boundary conditions, the
exponent of the leading correction corresponds to $\danomPf_{2,0}$
in \eqref{dimensionsI}, which is not the smallest one. Since this phase is
disordered and we do not break it explicitly in the computation of the
entanglement entropy, we expect only neutral fields under the $\mathbb{Z}_{r-2}$
symmetries to enter in \eqref{RentropyFixed1}. Thus, we find it natural to
interpret the leading correction in \eqref{RentropyFixed1} as due to the most
relevant thermal operator $\epsilon_1$, see \eqref{epsilonk}, which,
coincidentally, is also the gap-opening operator.
This interpretation is further corroborated by the observation that, expressing
\eqref{RentropyFixed1} in terms of the correlation length $\xi \simeq
\tilde{q}^{-1/2}$, the dimension of the leading correction becomes
$2\danomeps{1}$, which seems to be due to a spin-less operator.
Moreover, as we observed in the introduction of section \ref{entropyI}, by
changing $a$ we can generate only $\left[ r/2 \right]$ independent
combinations of primary fields and their characters, which coincides with the
number of allowed thermal operators in \eqref{epsilonk}.

\subsection{Full entropy}

We consider the full partition function, that is obtained with the dual transformation of the full sum in \eqref{partitionI}. One gets
\be
\label{partitionFunctionI}
\ZZ_\alpha =  \sum_{a=1}^{r-1} \left[ e^{- \frac{\ii \pi}{8}} \theta_1\left(\frac{a \pi}{r}, \ii \sqrt{p}\right)\right]^\alpha \ZZ_\alpha^{(a)} \; .
\ee
Also in this case, the series expansion of the $\theta_1$ function \eqref{thetadef}
is useful for integer $\alpha = n$
\bea
    \ln \mathcal Z_n & = &  - {\nupf  \over 12 n} \; \cpf \ln p +
{n \over 8} \ln p + \ln { 2^{1 +n} \over \sqrt{r}}
    \label{ZIsum} - \sum_{j=1}^\infty \ln \left( 1 - p^{ 2 r \, j \over (r-2)
n} \right)
    \nonumber \\
    && + \ln \sum_{j=0}^\infty p^{j(j+1)\over n (r - 2)} (-1)^j
    \sum_{a=1}^{r-1} \sin {\pi a (2j+1) \over r} \left[\sum_{k=0}^\infty (-1)^k
(-p)^{k(k+1)\over 2} \sin\left((2k+1) \frac{a\pi}{r}\right) \right]^n \; ,
    \nonumber
\eea
and again the partition function can be reproduced exactly with \eqref{sumdelta}
\be
\label{ZI1}
 \ZZ_1  = \sqrt r  \frac{e^{- \frac{\ii \pi}{8}} \; \theta_2 \left(0, \ii \sqrt{\tq}\right)}{\tq^{\frac{1}{12}} \prod_{j=1}^\infty (1 - \tq^{2 j})} \; .
\ee
For general $\alpha$ we can expand the Renyi entropy at desired order
\bea
 S_\alpha^{\rm (bulk)} & = & - \frac{\cpf}{24} \left(1 + \frac 1 \alpha\right)\log \tq
 + \frac{\log r}{2}+ \frac{1}{1-\alpha} \log\frac{2 s_\alpha(1,0)}{r}
 \nonumber \\
 && - \frac{1}{1-\alpha}\frac{s_\alpha(3,0)}{s_\alpha(1,0)} \tq^{2\over r\alpha} +
 \Ord{p^{4\over (r-2)\alpha}} \; ,
\label{RentropySum1}
\eea
where we see that the leading correction comes from the same
$\danomeps{1}$ operator as in \eqref{RentropyFixed1}. 

Thus, we see that, unlike for regime III, in the disordered phase the leading
correction is less sensitive to the boundary condition at the origin and
coincide with the scaling dimension of the gap-opening field.

\subsection{The boundary contribution}

Now we turn to the term related to the boundary heights $f_{bc}(q)$: from its definition in \eqref{boundaryI}, we see that for the set of values of $b,c$ that makes it non-trivial, it gives rise to a peculiar set of terms appearing in the expansion of the Renyi entropy
\be
\label{boundaryCorrections}
S_\alpha^{(bc)} = \frac{1}{1-\alpha} \ln \left[\frac{\epsilon_b \left(x^{\alpha(r-2)}\right)}{\epsilon_b \left(x^{r-2}\right)^\alpha}\right] =
\ln b + \frac{(b^2-1) \pi^4 \alpha}{
 6  (\ln \tq)^2} + O\left(\frac{1}{\ln \tq}\right)^4 \; .
\ee
We see that the boundary contribution modifies the constant term (boundary term \cite{affl1991}) and generates sub-leading logarithmic corrections.
This result may appear surprising: when local quantities are computed as in
\eqref{lhp}, such term cancels out, as expected since we are in a disordered
phase. But in the Renyi entropies, being a highly non-local object, also the
boundary appears. Of course, in considering the entropy of an actual state, one
might need to sum over different boundary conditions, possibly with different
weights, and the boundary contribution to the entropy might change
significantly. Thus, it might be pointless to try to provide a CFT
interpretation of \eqref{boundaryCorrections}, as these boundary conditions
might not have any conformal counterpart. However, it should be noted that, once
expressed in terms of the correlation length, the logarithmic corrections in
\eqref{boundaryCorrections} have the same form as those predicted in a CFT with
a (bulk) marginal field \cite{erik2008, card2010}. 
This is peculiar, since $\mathbb{Z}_{r-2}$ parafermions present such a marginal
field only for certain given values of $r$. One can check that a massless flow
in the theory, essentially due to a free boson, is present, for example, for the
series
\be
\label{freebosonminimal}
 r = m(m+1) \qquad \Rightarrow \qquad \cpf = 1 + \left(1 - \frac{6}{m(m+1)}\right)
\ee
where the central charge takes the form of a free boson plus a unitary minimal model.
Since, the logarithmic corrections in the entanglement entropy \eqref{boundaryCorrections} typically appear for every $r>5$, as we will see in the next section, these terms must have a different origin.


\section{Some examples}
\label{examples}

To better elucidate our results, it is instructive to specialize
and consider two particular
examples: the Ising model ($c=1/2$) and the 3-state
Potts model ($c=4/5$). In fact, within the RSOS, we have two possible
realizations of these models: one in Regime III (respectively $r = 4$ and $r=6$) as unitary
minimal models and one in Regime I ($r = 4$ and $r = 5$) as parafermions. The
comparison between the two realizations of the same theory can shed some light on
the two phases and the nature of the corrections.

\subsection{Ising model}

The Ising model is arguably the simplest CFT, since it consists of only three operators: $1, \sigma, \epsilon$.
Characters of the Ising model are known to arise in the study of the CTM's of eight-vertex models \cite{card1988} , which has the same Yang-Baxter algebra as the RSOS. Depending on the choice of parameters, the 1-D quantum system corresponding to the eight-vertex model is either an anisotropic $XY$ model in a transverse magnetic field, or an $XYZ$ chain in zero field. The entanglement entropies of both models have been calculated analytically in the thermodynamic, bipartite limit we are also considering here. The first has both an ordered and a disordered phase \cite{pesc2004, fran2006, fran2007}, while the $XYZ$ model presents only the ordered phase \cite{erco2010, erco2011, erco2012} . In the direct parameter $x$ of section \ref{RTtable}, the generalized partition function in the ordered phase has been observed to be proportional to the character of the spin operator, while in the disordered it is a combination of the identity and energy:
\bea
    {\cal Z}_\alpha^{\rm Ord} & \propto & \prod_{n=1}^\infty \left( 1 + x^{2n} \right)
    \quad = 2 \; x^{-1/12} \chi_\sigma \left( x^2 \right) \; ,
    \label{ZOrd} \\
    {\cal Z}_\alpha^{\rm Dis} & \propto & \prod_{n=1}^\infty \left( 1 + x^{2n-1} \right)
    = x^{1/24} \left[ \chi_1 \left( x^2 \right) + \chi_\epsilon \left( x^2 \right) \right] \; .
    \label{ZDis}
\eea

The Kac table for the Ising minimal and parafermionic model can be summarized as
\be
\label{isingCFT}
\left\{\begin{array}{lclclcl}
 \danom_1 & \equiv & \danom_{1,1} & = & \danomPf_{0,0} & = & 1 \; , \\
 \danom_\sigma & \equiv & \danom_{2,2} & = & \danomPf_{1,1} & = & \frac{1}{16} \; ,\\
 \danom_\epsilon & \equiv & \danom_{1,3} & = & \danomPf_{2,0} & = & \frac{1}{2} \; .
\end{array}\right.
\ee

In regime III, as explained in section \ref{minchar}, tuning the boundary conditions at the origin and at infinity, we generate each individual character. The entropy at fixed origin height then reflects the operator content of the theory under the duality transformation. Thus, looking at the modular matrix \eqref{modularmatrix}, we see that if we start with the identity or the energy field, the first correction to the entropy \eqref{RentropyFixed} comes from the most relevant operator, i.e. the spin operator $\sigma$. However, if the height at the origin is set to $a=2$, the coefficients of the $\sigma$ contributions vanishes (as the modular matrix has zero element for the $(\sigma,\sigma)$ entry) and the most relevant correction is given by the energy, as in \cite{fran2007, erco2012}.

As we explained in section \ref{fullentropyIII} for the general case, the $\danom_{2,2}$ field is odd under $\mathbb{Z}_2$ and therefore disappears in the full entropy \eqref{RentropySum}, and only the energy and identity appear. Of course, the field $ \danom_{3,3} $, which in general would give most relevant correction in \eqref{RentropySum}, does not appear in the Kac table of the Ising model and indeed its coefficients are vanishing.

In regime I, things are a bit different. By direct inspection of \eqref{thermoI}, we see that for $a=2$ \eqref{ZDis} is realized and both $a=1$ and $a=3$ give \eqref{ZOrd}. Thus, as we conjectured at the end of section \ref{centralspinI}, not every combination of operators and their characters appear. In this case it seems that fields with the same parity under $\mathbb{Z}_2$ appear together. It is then straightforward to see that after the modular transformation the character of the spin operator is never generated and the leading correction to the entropy is always given by the energy $\epsilon$, both when fixing the central height at any value as in \eqref{RentropyFixed1} and by summing over it \eqref{RentropySum1} .
Indeed, the identity and the energy are the fields in \eqref{epsilonk}.

Moreover, it is easy to check that the entropy contribution \eqref{boundaryCorrections} due to the boundary condition at infinity is always vanishing since for all the allowed values of $b,c$ in (\ref{boundaryI}, \ref{boundaryIb})
we have $f_{bc}(q) = 1$. 

As a final remark, we notice that for the Ising model, through the identities collected in the appendix and some manipulations, the partition functions (\ref{ZalphaIII}, \ref{partitionI}) can be written in a relatively explicit way. One simplification arises because, due to the $\mathbb{Z}_2$ symmetry, it is sufficient to fix the boundary conditions to $b=1, c=2$.  For regime III we have
\bea
   {\cal Z}_\alpha^{\rm (III)} & = & x^{\alpha \over 24} \Big[ \chi_1 \left( x^{2 \alpha} \right) + \chi_\epsilon \left( x^{2 \alpha} \right) \Big]
   \prod_{n=1}^\infty \left( 1 - x^n \right)^\alpha \left( 1 + x^{2n} \right)^\alpha
   \nonumber \\
   && + \; x^{-{\alpha \over 12}} \; \chi_\sigma \left( x^{2 \alpha} \right)
   \prod_{n=1}^\infty \left( 1 - x^n \right)^\alpha \left( 1 + x^{2n -1} \right)^\alpha ,
   \label{ZIsingIII}
\eea
and for regime I
\bea
   {\cal Z}_\alpha^{\rm (I)} & = & x^{{1 \over 12}\alpha} \Big[ \chi_1 \left( x^{4 \alpha} \right) + \chi_\epsilon \left( x^{4 \alpha} \right) \Big] x^{-\alpha}
   \prod_{n=1}^\infty \left( 1 - x^{2n} \right)^\alpha \left( 1 + x^{4n} \right)^\alpha
   \nonumber \\
   && + \; x^{-{\alpha \over 6}} \; \chi_\sigma \left( x^{4 \alpha} \right) \left[ 1 + (-1)^\alpha \right] x^{-\alpha}
   \prod_{n=1}^\infty \left( 1 - x^{2 n} \right)^\alpha \left( 1 + x^{4n -2} \right)^\alpha .
   \label{ZIsingI}
\eea
In regime I, the coefficients of the $a=1$ and $a=3$ terms are equal and opposite and we see that for $\alpha = 2m-1$ the partition function is simply proportional to the one found in the disordered phase of the $XY$ model \eqref{ZDis}, consistently with the fact that this regime is also disordered. We also notice that for $\alpha =1$, the coefficients in (\ref{ZIsingIII}, \ref{ZIsingI}) have the same form as (\ref{ZOrd}, \ref{ZDis}). Thus, the partition function can be formally written as a bilinear in the characters of the model. This reminds us of what observed in \cite{erco2012} and we take it as further indication that the character structure of the CTM in integrable models is mostly due to the analytical structure that permeates this beautiful construction, and not on some underlying Virasoro algebra.

\subsection{3-state Potts model}

The operator content of the minimal model in this case is given by:
\be
\label{pottsCFT}
\danom = \left\{0, \frac 1 {40}, \frac 1 {15}, \frac 1 8,
\frac 2 5, \frac {21} {40}, \frac 2 3, \frac 7 5, \frac {13} 8, 3\right\} \; .
\ee
The most relevant is $\danom_{2,2} = \frac 1 {40}$, which is the one appearing in
\eqref{RentropyFixed}, except for $a=3$. As usual, this field cannot enter in the full entropy obtained summing over the central height,
and the leading correction comes from the next relevant operator with $\danom_{3,3}= \frac 1 {15}$.

In the parafermionic realization, we only have four allowed conformal anomalies
\be
\label{pottsPF}
\danomPf = \left\{0, \frac 1 {15}, \frac 2 5, \frac 2 3 \right\} \; .
\ee
The leading correction in the entropies (\ref{RentropyFixed1}, \ref{RentropySum1}) is coming from $\danomeps{1}=\danomPf_{2,0} = \frac 2 5$, as we found in \eqref{RentropyFixed1}.
We notice that in this case, as it was for the Ising model, only two thermal operators \eqref{epsilonk} exist ($k=0,1$) 
and the boundary corrections again disappears for any allowed choice of
$b,c$.

We see that, contrary to the Ising example, here the two different (minimal ordered and parafermionic disordered) realizations of the 3-state Potts model have different corrections in the entanglement entropy.

The CTM's spectra of the 3- and 5-state Potts model have been calculated numerically in \cite{ritt1999} with a DMRG approach and an impressive agreement with the analytical expectations was found, also in the presence of integrability breaking terms, sufficiently close to criticality. This indicates that our results for the entanglement entropy should also remain valid under the same conditions.

It is worth to recall here that this parafermionic realization of the 3-state Potts model appears as the
ferromagnetic phase in the Fibonacci chain \cite{feig2007}. As stressed in
\cite{treb2008}, due to the topological symmetry present in the quantum
realization, all the relevant perturbations are forced to vanish, and the
critical point is topologically protected. It means that,
in the RSOS, the topological
symmetry is restored only at the gapless points. Therefore it would be interesting to compare our
predictions with the numerical data for the entanglement entropy coming from
the anyonic chain, once a perturbation breaking the topological charge is
turned on.

\section{Conclusions}
 \label{conclusions}

Following the suggestion put forward in \cite{cala2010}, we show how to calculated the bipartite Renyi entropy in the thermodynamic limit of a set of models known as RSOS. The method we employed is quite general and powerful and requires just the knowledge of the structure of the Corner Transfer Matrix eigenvalues of the system under consideration. In our case, the model being exactly solvable, the CTM spectrum is fully known analytically, thanks to \cite{andr1984}. However, generic systems close to criticality are expected to organize their CTM eigenvalues according to the CFT reached at criticality, as seen, for instance, in \cite{ritt1999}. If one was able to determine the coefficients in the expansion of the CTM in terms of characters of the CFT, the approach we used in this work would apply directly.

Beside it feasibility, this study of the RSOS was motivated by the fact that
this model provides a lattice realization of all minimal and parafermionic
conformal models. It is remarkable that a single system can realize such a variety of phase transitions and thus its entropy provides a unique case study for the approach to criticality in $1+1$ dimensions.
The different CFT's are realized by varying an integer parameter $r$, while the continuous parameter $p$ (or its dual $x$) measures the departure from criticality. Furthermore, the boundary conditions play an important role in fixing the phase under consideration.

We were thus able to compute the dependence of the Renyi entropy on $p$ and to study its behavior. The expansion of the entropy in regime III is given by \eqref{RentropyFixed}, if we project the Hilbert space on a subset specified by fixing the central height in the CTM, and by \eqref{RentropySum} for the absolute ground state. In regime I, we have \eqref{RentropyFixed1} for the projected case and \eqref{RentropySum1} for the whole case, with the addition of the boundary term \eqref{boundaryCorrections}, when present.

Our results confirm the expectation in \eqref{Sexp}, according to which,
approaching the critical points, we have a leading logarithmic term with a
universal prefactor (set by the conformal anomaly), a non universal constant term,
and power-law corrections with non-universal coefficients. We related the
exponents of the corrections to the conformal dimensions of one of the critical
fields. The leading correction always has the form of an {\it unusual
correction}, using the terminology of \cite{card2010}, and its dimension is that
of the most relevant field allowed. By changing the boundary
conditions on the RSOS, we can select different states for which we calculate
the bipartite Renyi entropy, and we noticed
that certain corrections can be suppressed and thus the leading term can be
determined by different operators.
In particular, we found that symmetry considerations prevent
the appearance of the most relevant field in the Renyi entropy of the the
absolute ground state. In the case of the minimal models, where the most
relevant field $\danom_{2,2}$ is the order parameter, the leading contribution
is given by the next most relevant operator, that is $\danom_{3,3}$. For
parafermionic model, the effect is even more dramatic, because the
$\mathbb{Z}_N$ symmetry seems to select only certain fields and the first
correction generally comes from the most relevant operator neutral under the symmetry, that is $\danomeps{1}$ in \eqref{epsilonk}.

In our opinion, these sort of effects due to the boundary conditions could represent an interesting possibility for numerical studies in this and other models, where the operator content of the theory can be in principle read out, by a proper turning of the boundary conditions.

In the parafermionic phase, we also observed the emergence of non power-law
corrections, of the same logarithmic form $(\log \xi)^{-2n}$ expected in the presence of a primary
marginal field in the theory. These types of terms were already found in \cite{erco2012}
and would be in agreement with a na\"ive scaling argument applied to the $(\log
\ell)^{-2}$ terms of \cite{card2010}, where the expansion is computed for a finite interval of length $\ell$. 
However, we already pointed out that these logarithmic corrections are present even when the parafermionic theory does not support a marginal field, and thus we should conclude that the origin of these terms is not so simple and might be a lattice effect due to non-conformal boundary conditions.

Another possible interpretation is that in general the relation between the corner transfer matrices and the Virasoro characters is ``accidental'', in the sense that is purely due to the analytical structure of both quantities. Both are elliptic functions: the latter bi-periodic in real space, while the former in parameter space (we remind that the elliptic nome $q$ has a different physical interpretation in the two cases).  When expanded close to the critical point, for consistency the CTM has to give the correct central charge of the gapless CFT, and this constraints the structure of the elliptic series defining the CTM. Since the same constraint applies to the Virasoro characters, this might explain why in general one can write the CTM as a sum of characters and why in the RSOS we did not find any connection between the dimension of the operator opening the gap and the dimension of the most relevant correction in the Renyi entropy. And it might explain why, playing with the boundary condition, one can 
turn on logarithmic corrections with no counterpart in the CFT.

Finally, let us remark that the original work \cite{andr1984} on the RSOS spent a considerable effort in developing advanced mathematical identities (known has generalized Rogers-Ramanujan identities) to access the partition functions of the model. In our calculations, we overcome some difficulties involved with summing up Gaussian polynomials, by performing first a duality transformation that, in our cases, turned a product of Gaussian polynomials into a sum over exponential one, which are easy to handle. We do not know how general and applicable this approach is, but it revealed to be quite powerful for us.

\section*{Acknowledgments}

We wish to thank Paul Pearce, Ingo Peschel, Giuseppe Mussardo, Jacopo Viti, Paul Fendley, Elisa Ercolessi, Francesco Ravanini for useful and very pleasant discussions.
This work was partially supported by the U.S. Department of Energy under cooperative research Contract Number DE-FG02-05ER41360. FF was supported by a Marie Curie International Outgoing Fellowship within the 7th European Community Framework Programme (FP7/2007-2013) under the grant PIOF-PHY-276093.

\appendix
\section{Elliptic functions and q-series}
\label{ellapp}

In this appendix we recall standard definitions and useful identities for
elliptic functions and $q$-series, used in the derivations in the text. For a
more detailed treatment and for the derivations of the various equalities, we
refer the reader to one of the standard textbooks on the topic, e.g. \cite{lawd1989, apos1989}.

First of all, the Jacobi Elliptic $\theta$ functions are defined as
\begin{subequations}
\begin{align}
 \theta_1(z,q) &= 2 \sum_{n = 0}^\infty (-1)^n q^{(n + 1/2)^2} \sin [(2n + 1)
z] \; ,\\ \label{theta1}
 \theta_2(z,q) &= 2 \sum_{n = 0}^\infty q^{(n + 1/2)^2} \cos[(2n+1)z] \; ,\\
 \theta_3(z,q) &=  1 + 2 \sum_{n=1}^\infty q^{n^2} \cos (2 n z) \; , \\
 \theta_4(z,q) &=  1 + 2 \sum_{n=1}^\infty (-1)^n q^{n^2} \cos (2 n z) \; .
\end{align}
\label{thetadef}
\end{subequations}

Employing the Jacobi triple product identity
\be
\label{jacobitriple}
\pochh{x^2}{x^2} \pochh{x y^2}{x^2} \pochh{x y^{-2}}{x^2} = \sum_{n =
-\infty}^{\infty} x^{n^2} y^{2n} \; ,
\ee
where we introduced the \textit{q-Pochhammer symbol}
\begin{align}
\label{qpochh}
\pochh{a}{q} &= \prod_{k = 0}^\infty \left( 1 - a q^k \right) \; , \\
(q)_\infty &= \pochh{q}{q} \; ,
\label{qinf}
\end{align}
one can derive the product representations for the $\theta$ functions:
\begin{subequations}
\begin{align}
 \theta_1(z,q) &= 2  \pochh{q^2}{q^2} q^{\frac 1 4} \sin z \prod_{n=1}^\infty
[1 - 2q^{2n} \cos(2z) + q^{4n}] \; , \\
 \theta_2(z,q) &= 2  \pochh{q^2}{q^2} q^{\frac 1 4} \cos z \prod_{n=1}^\infty
[1 + 2q^{2n} \cos(2z) + q^{4n}] \; ,\\
 \theta_3(z,q) &=  \pochh{q^2}{q^2} \prod_{n=1}^\infty
[1 + 2q^{2n-1} \cos(2z) + q^{4n-2}] \; ,\\
 \theta_4(z,q) &=  \pochh{q^2}{q^2} \prod_{n=1}^\infty
[1 - 2q^{2n-1} \cos(2z) + q^{4n-2}] \; .
\end{align}
\label{thetaprod}
\end{subequations}

In the text we also used the function
\be
   E(z,x) \equiv
   \pochh{z}{x}\pochh{x z^{-1}}{x} \pochh{x}{x}
   = \sum_{n=-\infty}^\infty (-1)^n x^{\frac{n(n-1)}{2}}z^n \; ,
   \label{Edef}
\ee
where the second equality follows again from \eqref{jacobitriple}.

The duality transformation for $\theta$ functions can be derived using the
Poisson summation formula, obtaining the so called \textit{Jacobi identities}.
Once we define $q$ and $\tq$ such that
\be
 q = e^{\ii \pi \tau} \; ,\qquad \qquad
 \tq = e^{-\frac{\ii \pi}{\tau}} \; , \qquad \qquad
 \Im \tau > 0 \; ,
\ee
they take the form
\begin{subequations}
\begin{align}
 \theta_1(z,\tq) &= -\ii(\ii \tau)^{\frac 1 2} e^{\frac{\ii \tau z^2}{\pi}}
\theta_1( \tau z, q) \\
 \theta_2 (z , \tq) &= (-\ii \tau)^{\frac 1 2} e^{\ii \tau z^2 \over \pi}
\theta_4 (\tau z, q) \\
\theta_3 (z , \tq) &= (-\ii \tau)^{\frac 1 2} e^{\ii \tau z^2 \over \pi}
\theta_3 (\tau z, q) \\
\theta_4 (z, \tq) &= (-\ii \tau)^{\frac 1 2} e^{\ii \tau z^2 \over \pi} \theta_2
(\tau z, q)
\end{align}
\label{jacobiIdentities}
\end{subequations}
Finally, it is possible to rexpress the function $E(z,x)$ in
\eqref{Edef} by means of the $\theta$ functions
\bea
    E  \left( \eu^{2 \ii z}, q^2 \right) & =  \ii q^{-1/4} \eu^{\ii z} \theta_1
(z, q) \; , \\
    E  \left( - \eu^{2 \ii z}, q^2 \right) & =  q^{-1/4} \eu^{\ii z} \theta_2
(z, q) ,
\label{EwithTheta}
\eea
Combining these expression with \eqref{jacobiIdentities}, it is possible to
obtain the expression of the partition functions in the dual variables.
\bibliographystyle{apsrev}
\bibliography{references}
\end{document}